\documentclass[aps,prd,amssymb,nofootinbib,twocolumn,epsf,showpacs]{revtex4}
\usepackage[usenames]{color} 
\usepackage{ulem} 
\usepackage{graphicx}
\usepackage{amssymb}
\usepackage{amsmath}
\usepackage{epstopdf}

\begin{document}


\title{A Spectral Approach to the Relativistic Inverse Stellar 
Structure Problem}

\author{Lee Lindblom and Nathaniel M. Indik}

\affiliation{Theoretical Astrophysics 350-17, California Institute of
Technology, Pasadena, CA 91125}

\date{\today}

\begin{abstract}
A new method for solving the relativistic inverse stellar structure
problem is presented.  This method determines a spectral
representation of the unknown high density portion of the stellar
equation of state from a knowledge of the total masses $M$ and radii
$R$ of the stars.  Spectral representations of the equation of state
are very efficient, generally requiring only a few spectral parameters
to achieve good accuracy. This new method is able, therefore, to
determine the high density equation of state quite accurately from
only a few accurately measured $[M,R]$ data points.  This method
is tested here by determining the equations of state from mock
$[M,R]$ data computed from tabulated ``realistic'' neutron-star
equations of state.  The spectral equations of state obtained from
these mock data are shown to agree on average with the originals to
within a few percent (over the entire high density range of the
neutron-star interior) using only two $[M,R]$ data points. Higher
accuracies are achieved when more data are used.  The accuracies of
the equations of state determined in these examples are shown to be
nearly optimal, in the sense that their errors are comparable to the
errors of the best-fit spectral representations of these realistic
equations of state.
\end{abstract}


\maketitle


\section{Introduction}
\label{s:introduction}

The standard stellar structure problem consists of determining the
structure of a star by solving the coupled gravitational and
hydrodynamic equations with an assumed equation of state for the
stellar matter.  The solutions to the standard problem determine the
various observable properties of the stars with a given equation of
state like their total masss $M$, their total radii $R$, etc.  The
inverse stellar structure problem determines what equation of state is
required to produce stellar models having a given set of macroscopic
obserables.  The goal of this paper is to find efficient and robust
methods of solving this inverse stellar structure problem.  The method
developed here is based on the use of spectral expansions to represent
the equation of state.  The values of the spectral coefficients in
these expansions are fixed in this method by matching stellar models
based on these equations of state to observed values of the masses and
radii of the stars.  Once fixed, these coefficients determine the
equation of state that represents the (approximate) solution to the
inverse stellar structure problem.

For non-rotating stars in general relativity theory, the simplest
version of the stellar structure equations were first derived by
Oppenheimer and Volkoff~\cite{Oppenheimer1939},
\begin{eqnarray}
\frac{dm}{dr} &=& 4\pi r^2\epsilon,
\label{e:OVm}\\
\frac{dp}{dr} &=& - (\epsilon+p)\frac{m+4\pi r^3 p}{r(r-2m)},
\label{e:OVp}
\end{eqnarray}
where $m(r)$ represents the mass contained within a sphere of radius
$r$; $p(r)$ is the pressure; and $\epsilon(p)$ is the total energy
density of the fluid.  Solving these equations with a given equation
of state is the standard relativistic stellar structure problem.  Once
an equation of state $\epsilon=\epsilon(p)$ is specified, these
equations determine a one parameter family of stellar models,
$m=m(r,p_c)$ and $p=p(r,p_c)$, where $p_c$ is the value of the
pressure at the center of the star $r=0$.  These solutions then
determine various macroscopic properties of the stars, including their
outer radii $R(p_c)$ where $p[R(p_c),p_c]=0$, and their total masses
$M(p_c)=m[R(p_c),p_c]$.  These macroscopic properties are (at least in
principle) observable.

The standard stellar structure problem can be thought of as a map that
takes the equation of state [a curve in energy density -- pressure
  space $\epsilon=\epsilon(p)$], into a curve in the space of
macroscopic observables, e.g. $[M(p_c),R(p_c)]$.  The inverse stellar
structure problem consists of finding the inverse of this
map~\cite{Lindblom1992}, i.e. determining the equation of state of the
stellar matter from a knowledge of some information about the
macroscopic structures of the stars (like their masses $M$ and radii
$R$).  The solution to this problem, like the solutions to many inverse
problems, is less straightforward than the solution to the standard
problem.

The inverse stellar structure problem is probably more relevant for
practical relativistic astrophysics, however, than the standard
problem.  The highest density part of the equation of state in neutron
stars, for example, is not well known.  Matter in this state is well
beyond the reach of laboratory experiments, so there is no independent
way of directly measuring its properties, including its equation of
state.  Numerous attempts have been made to model this matter
theoretically, but even today there is no consensus among
theoreticians.  Predictions of the energy density for a typical
neutron-star central pressure, for example, often differ by an order
of magnitude.  Therefore, the standard stellar structure problem for
neutron stars is not terribly useful.  In contrast, the inverse
problem may provide an important tool for learning about high density
nuclear matter.  Numerous high quality measurements of the masses of
neutron stars are now available~\cite{Lattimer2007}, and a few (fairly
imprecise and model-dependent) radius measurements are starting to
become available as well~\cite{Steiner2010, Galloway2012, Guver2012a,
  Guver2012b}.  In principle then, the inverse stellar structure
problem should (eventually) allow us to measure the high density
equation of state of neutron-star matter directly.  This measurement
would provide important information about nuclear interactions that
can not be obtained at present in any other way.

One naive approach to solving the inverse stellar structure problem
for neutron stars would simply be to match their observed properties,
e.g. their $[M,R]$ data, with models of those stars based on different
micro-physical models of the dense material in their cores.  In this
approach the model equation of state whose stellar models best matches
the data would be declared the observed neutron-star equation of
state.  This approach would clearly be ideal if there were wide
consensus on exactly what the high density core material is, and if
there were a reasonably simple model for this material that were
known, up to a few undetermined parameters that could be fixed by
these observations.  Unfortunately the wide diversity of ``realistic''
neutron-star equations of state in the literature, suggest that (in
the near term at least) this approach is not likely to be effective or
conclusive.  

A more practical variation of this approach uses some knowledge about
the expected properties of neutron-star matter in an intermediate
range of densities, and a more empirical description of the equation
of state for larger densities~\cite{Steiner2010, Steiner2012}.  This
approach is more promising, but the proposed model equations of state
of this type have many free parameters that must all be fit by the
observational data.  Since these data are likely to be sparse for some
time, we take a different approach here.  

Our goal is to find efficient and robust methods for solving the
inverse stellar structure problem that use no prior knowledge of the
high density micro-physics at all.  A (somewhat impractical) method
for solving the inverse stellar structure problem that uses no
information about the micro-physics of the high density equation of
state was given in the literature about 20 years
ago~\cite{Lindblom1992}.  This traditional method can be summarized as
follows. The total masses $M$ and radii $R$ of all of the stars
associated with a particular equation of state are assumed to be
known. The equation of state is also assumed to be known up to some
pressure $p_i$ with corresponding energy density
$\epsilon_i=\epsilon(p_i)$.  Let $M_i=M(p_i)$ and $R_i=R(p_i)$ denote
the mass and radius of the star whose central pressure is $p_c=p_i$.
Now choose another point, $[M_{i+1},R_{i+1}]$, along the mass-radius
curve, with a slightly larger central pressure.  The outer layers of
this new star are composed of low pressure material, $p\leq p_i$,
where the equation of state is known.  The stellar structure
equations, (\ref{e:OVm}) and (\ref{e:OVp}), can therefore be solved in
this outer region starting at the surface of the star, $r=R_{i+1}$,
where $p(R_{i+1})=0$ and $m(R_{i+1})=M_{i+1}$, by integrating inward
toward $r=0$.  This integration can be continued until the point
$r=r_{i+1}$ where $p(r_{i+1})=p_i$ and the known equation of state
ends.  This integration determines the radius $r_{i+1}$ and the mass
$m_{i+1}=m(r_{i+1})$ of a ``small'' core of high pressure material,
$p\ge p_i$ where the equation of state is not yet known.  If this core
is small enough, the stellar structure equations can be solved there
as a power series expansion about $r=0$.  The coefficients in this
expansion are functions of the central density $\epsilon_{i+1}$ and
pressure $p_{i+1}$ of this little core.  Since the mass and radius of
this core are known, $m_{i+1}$ and $r_{i+1}$, this power series can be
``inverted'' to determine $\epsilon_{i+1}$ and
$p_{i+1}$~\cite{Lindblom1992}.  This new point
$[\epsilon_{i+1},p_{i+1}]$ provides a small extension of the equation
of state beyond $[\epsilon_i,p_i]$.  Iterating these steps then
determines a sequence of closely-spaced points along the high density
portion of the equation of state curve.

This traditional solution to the inverse stellar structure problem is
unfortunately very impractical.  A large number of points $[M_i,R_i]$
are needed from the mass-radius curve to achieve modest accuracy in
the calculation of the corresponding points $[\epsilon_i,p_i]$ along
the equation of state curve.  Since $[M_i,R_i]$ points are very
difficult to measure (at least for neutron stars) it is unlikely that
there will ever be enough data to use this traditional method to
determine the unknown high density part of the neutron-star equation
of state.

This paper proposes a rather different approach to the solution of the
inverse stellar structure problem, an approach that can be very
effective even when only a small number of $[M_i,R_i]$ data points are
available.  The equation of state in this new approach is expressed as
a parametric equation, e.g. $\epsilon=\epsilon(p,\gamma_k)$, instead
of a table of values $[\epsilon_i,p_i]$ .  The parameters $\gamma_k$
are adjusted to give the best-fit approximation to a particular
equation of state model.  Parametric representations of this sort,
based on spectral expansions, have been shown to be extremely
efficient at representing the high density portions of ``realistic''
neutron-star equations of state~\cite{Lindblom10}.  Only a few
non-vanishing $\gamma_k$ are generally needed to achieve 1\% accuracy
in most cases.

The basic idea of this new method for solving the inverse stellar
structure problem is to choose the equation of state parameters
$\gamma_k$ by minimizing the differences between the masses and radii
of real neutron stars, $M_i$ and $R_i$, with those based on the
parametric model equation of state, $M(p_c,\gamma_k)$ and
$R(p_c,\gamma_k)$.  Once the $\gamma_k$ are fixed, the parametric
equation $\epsilon=\epsilon(p,\gamma_k)$ then provides an approximate
solution of the inverse stellar structure problem.  Spectral
expansions typically converge exponentially as the number of terms in
the expansion are increased (for smooth functions).  These approximate
solutions to the inverse stellar structure problem are therefore
expected to converge to the exact equation of state as the number of
data points $[M_i,R_i]$ and the number of parameters $\gamma_k$ fixed
by this method are increased.

The remainder of this paper presents details on how to implement this
new spectral approach to the solution of the inverse stellar structure
problem, along with practical tests of its accuracy and efficiency.
Section~\ref{s:SpectralEOS} reviews the particular spectral
representation of the equation of state used in the solution presented
here.  Section~\ref{s:FixingSpectralParameter} describes how the
spectral parameters $\gamma_k$ are fixed by matching to the given
$[M_i,R_i]$ data points.
Section~\ref{s:TestingSpectralInversionMethod} presents a series of
numerical tests of the accuracy and efficiency of this new method.
Mock $[M_i,R_i]$ data computed from a collection of 34 ``realistic''
neutron-star equations of state are used as input in these tests.
These tests show, for example, that the resulting spectral equation of
state agrees with the exact to within a few percent (on average) when
only two $[M_i,R_i]$ data points are used.  Higher accuracies are
(generally) achieved when more data points are used.
Section~\ref{s:Discussion} discusses some of the limitations of the
numerical tests presented here, and proposes several ways that the
basic method developed here might be extended and improved.  Some of
the more complicated technical details needed to implement this method
are described in two Appendices.  Appendix~\ref{s:AppendixA} describes
how to evaluate the derivatives of $M(h_c,\gamma_k)$ and
$R(h_c,\gamma_k)$, with respect to the parameters $h_c$ and
$\gamma_k$.  Appendix~\ref{s:AppendixB} describes the interpolation
method used here to bridge the gaps between points in the exact
``realistic'' equation of state tables.


\section{Spectral Representations of the
Equation of State}
\label{s:SpectralEOS}

The version of the relativistic stellar structure equations most
useful for our analysis here requires that the equation of state be
written in a form where the energy density $\epsilon$ and pressure $p$
are given as functions of the relativistic enthalpy, $h$.  The usual
form of the equation of state, $\epsilon=\epsilon(p)$, must therefore
be re-written as a pair of equations $\epsilon=\epsilon(h)$ and
$p=p(h)$, where the enthalpy $h$ is defined as
\begin{eqnarray}
h(p) = \int_0^p \frac{dp'}{\epsilon(p') + p'}.
\label{e:EnthalpyDef}
\end{eqnarray} 
The needed expressions, $\epsilon=\epsilon(h)$ and $p=p(h)$, can be
constructed by inverting $h=h(p)$ from Eq.~(\ref{e:EnthalpyDef}) to
obtain $p=p(h)$, and then by composing the result with the standard
form of the equation of state, $\epsilon=\epsilon(p)$, to obtain
$\epsilon(h)=\epsilon[p(h)]$.

The transformations needed to construct $\epsilon=\epsilon(h)$ and
$p=p(h)$ in this way are difficult to perform efficiently and
accurately in numerical computations.  Therefore it is best to
construct a spectral representation of the equation of state that is
based directly on $h$.  This can be done by introducing an enthalpy
based spectral expansion of the adiabatic index $\Gamma$~\cite{Lindblom10}:
\begin{eqnarray}
\Gamma(h) &\equiv& \frac{\epsilon + p}{p}\frac{dp}{dh}
\left(\frac{d\epsilon}{dh}\right)^{-1},
 \label{e:GammaDef}\\  
&=& \exp\left\{\sum_k \gamma_k 
\left[\log\left(\frac{h}{h_0}\right)\right]^k\right\},
\label{e:Gamma}
\end{eqnarray}
where $h_0$ is the lower bound on the enthalpy, $h_0\leq h$, in the
domain where the spectral expansion is to be used.  This is a standard
spectral expansion of the function $\log \Gamma(h)$ in which the
$[\log(h/h_0)]^k$ are the spectral basis functions and the 
$\gamma_k$ are the spectral expansion coefficients (or parameters).

The functions $p(h)$ and $\epsilon(h)$ are obtained from $\Gamma(h)$
by integrating the system of ordinary differential equations,
\begin{eqnarray}
\frac{dp}{dh} &=& \epsilon + p,\label{e:dpdh}
\\
\frac{d\epsilon}{dh} &=& \frac{(\epsilon + p)^2}{p  \Gamma(h)},
\label{e:depsilondh}
\end{eqnarray}
that follow from the definitions of $h$ and $\Gamma$ in
Eqs.~(\ref{e:EnthalpyDef}) and (\ref{e:GammaDef}).  The general
solution to these equations can be reduced to quadratures:
\begin{eqnarray}
p(h)&=&p_0 \exp\left[\int_{h_0}^h \frac{e^{h'}dh'}{\mu(h')}
\right],\label{e:PressueH}\\
\epsilon(h)&=&p(h)  \frac{e^h -\mu(h)}{\mu(h)},
\label{e:EnthalpyH}
\end{eqnarray}
where $\mu(h)$ is defined as.
\begin{eqnarray}
\mu(h) = \frac{p_0\, e^{h_0}}{\epsilon_0  + p_0} 
+ \int_{h_0}^h \frac{\Gamma(h')-1}{\Gamma(h')} e^{h'}dh'.
\label{e:TildeMuDef}
\end{eqnarray}
The constants $p_0$ and $\epsilon_0$ are defined by $p_0=p(h_0)$ and
$\epsilon_0=\epsilon(h_0)$ respectively.  While these quadratures can
not be done analytically for the spectral expansion of $\Gamma(h)$
given in Eq.~(\ref{e:Gamma}), they can be done numerically very
efficiently and accurately using Gaussian quadratures.\footnote{The
  numerical accuracy (and hence the efficiency) of the numerical
  integrations in Eqs.~(\ref{e:PressueH}) and (\ref{e:TildeMuDef}) can
  be improved significantly by changing integration variables from $h$
  to $x=\log(h/h_0)$ before performing standard Gaussian quadratures.
  Our tests achieved accuracies about $10^{-11}$ for
  $\epsilon(h,\gamma_k)$ and $p(h,\gamma_k)$ with 10 Gaussian
  integration points using the $x$ variable, compared to about
  $10^{-3}$ accuracies for the same tests using the $h$ variable.}

The method of solving the inverse stellar structure problem proposed
in Sec.~\ref{s:FixingSpectralParameter} is based on this spectral
representation of the equation of state:
$\epsilon=\epsilon(h,\gamma_k)$ and $p=p(h,\gamma_k)$.  Any equation
of state can be represented approximately in this way by using a
finite number of spectral parameters $\gamma_k$ in the expression,
Eq.~(\ref{e:Gamma}), for $\Gamma(h)$.  In analogy with other spectral
expansions, the accuracy of these approximations are expected to
converge exponentially (for smooth equations of state) as the number
of spectral coefficients is increased.  Numerical tests that fit
``realistic'' neutron-star equations of state using this
method~\cite{Lindblom10} are consistent with this expectation about
the convergence of these expansions.


\section{Fixing the Spectral Parameters}
\label{s:FixingSpectralParameter}

The spectral approach to the inverse stellar structure problem fixes
the equation of state parameters $\gamma_k$ by choosing the values
whose stellar models best match a collection of points $[M_i,R_i]$
from the exact mass-radius curve.  The process of fixing the
$\gamma_k$ in this way can be made more efficient numerically by using
a somewhat non-standard version of the stellar structure equations.
Therefore, we digress briefly here to review this alternate
formulation.

The Oppenheimer-Volkoff version of the stellar structure problem,
Eqs.~(\ref{e:OVm}) and (\ref{e:OVp}), determines $m$ and $p$ as
functions of $r$.  That traditional approach has two inconvenient
features: First the integration domain, $[0,R]$, is only known after
the fact when the surface of the star at $r=R$ is found numerically by
solving $p(R)=0$.  Second, the equation $p(R)=0$ that defines the
surface of the star is somewhat difficult to solve numerically because
$dp/dr$ typically vanishes at $r=R$.  These inconveniences can be
avoided by transforming the equations into a form where $m$ and $r$
are determined as functions of the relativistic enthalpy $h$ (see
Ref.~\cite{Lindblom1992}):
\begin{eqnarray}
\frac{dm}{dh}&=&
-\frac{4\pi r^3 \epsilon(r-2m)}{m+4\pi r^3 p},
\label{e:OVEquation_m}\\
\frac{dr}{dh}&=&
-\frac{r(r-2m)}{m+4\pi r^3 p}
\label{e:OVEquation_r}.
\end{eqnarray}
Solving the equations in this form begins by specifying ``boundary''
conditions, $m(h_c)=r(h_c)=0$, at the center of the star where
$h=h_c$, and then integrating toward the surface of the star where
$h=0$.  The derivative $dr/dh$ in Eq.~(\ref{e:OVEquation_r}) is
non-zero and bounded at the surface of the star, so this formulation
completely eliminates the problems associated with solving $p(R)=0$ to
locate the star's surface.  This version of the problem is also easier
to implement numerically because it is carried out on the domain
$[h_c,0]$, which is fixed before the integration is performed.  The
total mass and radius of the stellar model are obtained in this
formulation simply by evaluating the solutions $m(h)$ and $r(h)$ at
the surface of the star where $h=0$: $M=m(0)$ and $R=r(0)$.  More
details about how to implement this formulation of the stellar
structure problem are given in Ref.~\cite{Lindblom1992} and in
Appendix~\ref{s:AppendixA} of this paper.

This alternate formulation of the stellar structure problem,
Eqs.~(\ref{e:OVEquation_m}) and (\ref{e:OVEquation_r}), requires that
the equation of state be expressed in terms of the enthalpy, i.e. that
$\epsilon=\epsilon(h)$ and $p=p(h)$ be provided.  The spectral
representations, $\epsilon=\epsilon(h,\gamma_k)$ and
$p=p(h,\gamma_k)$, described in Sec.~\ref{s:SpectralEOS} are therefore
ideal for this purpose.  The general solution to this form of the
stellar structure problem is a pair of functions of the form
$m(h,h_c,\gamma_k)$ and $r(h,h_c,\gamma_k)$.  These solutions are
determined uniquely by the parameter $h_c$, the central enthalpy of
the star, and $\gamma_k$, the spectral parameters that determine the
equation of state.  The total mass $M(h_c,\gamma_k)$ and radius
$R(h_c,\gamma_k)$ associated with one of these stellar models are
determined from these solutions by $M(h_c,\gamma_k)=m(0,h_c,\gamma_k)$
and $R(h_c,\gamma_k)=r(0,h_c,\gamma_k)$.

The new method of solving the inverse stellar structure problem, which
we introduce here, fixes the values of the spectral parameters
$\gamma_k$ by minimizing the differences between the model mass-radius
values $[M(h_c,\gamma_k),R(h_c,\gamma_k)]$ and points $[M_i,R_i]$ from
an exact mass-radius curve.  Thus we fix the values of the $\gamma_k$
by minimizing
\begin{eqnarray}
\chi^2(h_c^j,\gamma_k) &=& \frac{1}{N_\mathrm{stars}}\sum_{i=1}^{N_\mathrm{stars}}
\left\{\left[\log\left(\frac{M(h_c^i,\gamma_k)}{M_i}\right)\right]^2\right.
\nonumber\\
&&\qquad\qquad
+\left.\left[\log\left(\frac{R(h_c^i,\gamma_k)}{R_i}\right)\right]^2\right\},
\qquad
\label{e:Chi2Def}
\end{eqnarray}
with respect to each of the $\gamma_k$.  Each of the
$N_\mathrm{stars}$ stellar models used in this fit has a central
enthalpy, $h_c^i$, whose value is also needed (along with the
$\gamma_k$) to construct the mass-radius values
$[M(h_c^i,\gamma_k),R(h_c^i,\gamma_k)]$.  Since the $h_c^i$ will not be
known {\it a priori}, these additional parameters must also be
determined as part of the fitting process.  These parameters are
fixed, therefore, by minimizing $\chi^2(h_c^j,\gamma_k)$ with respect
to variations in each of the $h_c^j$.  

In summary then, this new method of solving the inverse stellar
structure problem determines the equation of state by fixing the
spectral parameters $\gamma_k$ in a way that minimizes the differences
between the model mass-radius values
$[M(h_c^i,\gamma_k),R(h_c^i,\gamma_k)]$ and values $[M_i,R_i]$ from an
exact mass-radius curve.  This minimization problem is equivalent to
solving the $N_\mathrm{stars}$ equations
\begin{eqnarray}
\frac{\partial\chi^2}{\partial h_c^i} =  0,
\label{e:MinimizeChi2hc}
\end{eqnarray}
and the $N_{\gamma_k}$ (the number of spectral parameters) equations
\begin{eqnarray}
\frac{\partial\chi^2}{\partial \gamma_k} &=& 0,
\label{e:MinimizeChi2gammak}
\end{eqnarray}
for the parameters $h_c^i$ and $\gamma_k$.  Since the number of
independent $M_i$ and $R_i$ data values used in this fitting process
is $2N_\mathrm{stars}$, it follows that the maximum number of spectral
parameters that can be fixed in this way is $N_{\gamma_k}\leq
N_\mathrm{stars}$.

A number of numerical methods for solving non-linear least squares
problems such as Eqs.~(\ref{e:MinimizeChi2hc}) and
(\ref{e:MinimizeChi2gammak}) are discussed in the literature.  Many of
these methods require only that $\chi^2(h_c^i,\gamma_k)$ be provided
numerically for arbitrary values of the parameters $h_c^i$ and
$\gamma_k$.  Some methods also require in addition that the values of
the derivatives $\partial\chi^2/\partial h_c^i$ and
$\partial\chi^2/\partial \gamma_k$ be provided.  The numerical tests
described in Sec.~\ref{s:TestingSpectralInversionMethod} use the
Levenberg-Marquardt method for solving these equations, and this
method requires that both the values and the derivatives of $\chi^2$
be provided.

The derivatives of $\chi^2$ are determined by the derivatives of
$M(h_c^i,\gamma_k)$ and $R(h_c^i,\gamma_k)$ with respect to $h_c^i$
and $\gamma_k$.  These derivatives can be approximated numerically by
expressions of the form, $\partial M/\partial\gamma_k \approx
[M(h_c^i,\gamma_k+\delta\gamma_k)
  -M(h_c^i,\gamma_k-\delta\gamma_k)]/2\delta\gamma_k$.  We find it is
more efficient (and more accurate) however to evaluate these
derivatives by solving an auxiliary system of ordinary differential
equations, that are obtained by differentiating
Eqs.~(\ref{e:OVEquation_m}) and (\ref{e:OVEquation_r}) with respect to
these parameters.  This method of evaluating the needed derivatives of
$M(h_c^i,\gamma_k)$ and $R(h_c^i,\gamma_k)$ is described in some
detail in Appendix~\ref{s:AppendixA}.


\section{Testing the Spectral Inversion Method}
\label{s:TestingSpectralInversionMethod}

In this section we test the spectral method of solving the inverse
stellar structure problem by analyzing sets of mock $[M_i,R_i]$ data points
from mass-radius curves based on known ``realistic'' neutron-star
equations of state.  We use these mock $[M_i,R_i]$ data to construct
best-fit values for the spectral equation of state parameters
$\gamma_k$, using the least-squares method outlined in
Sec.~\ref{s:FixingSpectralParameter}.  Then we compare the equation of
state $\epsilon(h,\gamma_k)$ constructed in this way with the exact
equation of state $\epsilon(h)$ used to compute the mock $[M_i,R_i]$ data
points.  We perform these comparisons using different numbers of
$[M_i,R_i]$ data points to determine how the accuracy of the
approximate equation of state improves as the number of data points is
increased.  We construct the mock $[M_i,R_i]$ data points using 34
different realistic neutron-star equations of state to explore how
well the method works for a fairly wide variety of different equations
of state.
 
The 34 equations of state used to construct the $[M_i,R_i]$ data
points used in these tests are the same ones used by Read, Lackey,
Owen and Friedman~\cite{Read:2008iy} in their study of approximate
piecewise polytropic fits to the equation of state.  These 34
realistic equations of state are based on a variety of different
models for the composition of neutron-star matter, and a variety of
different models for the interactions between the particle species
present in the model material.  Descriptions of these realistic
equation of state models, and references to the original publications
on each of these equations of state are given in
Ref.~\cite{Read:2008iy}, and are not repeated here.  The individual
equations of state are referred to here using the abbreviations used
in Ref.~\cite{Read:2008iy}, {\it e.g.}  PAL6, APR3, BGN1H1, etc.  The
list of these equations of state are given in the first column of
Table III of Ref.~\cite{Read:2008iy} as well as the first column of
Table~\ref{t:TableI} here.  Spectral fits have already been shown to
provide excellent approximations to these 34 equations of state in
Ref.~\cite{Lindblom10}.  Only two or three spectral coefficients
$\gamma_k$ are needed to achieve accuracies at the few percent level.
The new question being studied here, therefore, is whether the
spectral parameters $\gamma_k$ can be determined by fitting
$[M_i,R_i]$ data instead of fitting directly to the equation of state
itself.

These 34 exact realistic equations of state are provided to us as tables
of energy density and pressure points $[\epsilon_i,p_i]$.  We compute
the enthalpy values $h_i$ corresponding to the points in these tables,
interpolate between these tabulated points whenever necessary, and
construct complete enthalpy based equations of state
$[\epsilon(h),p(h)]$ from these tabulated data using the methods
described in Appendix \ref{s:AppendixB}.  Whenever we refer to one of
these exact realistic equations of state, we mean the one constructed
by interpolating the tabulated equation of state data in this way.

\begin{figure}[htbp!]
\centerline{\includegraphics[width=3in]{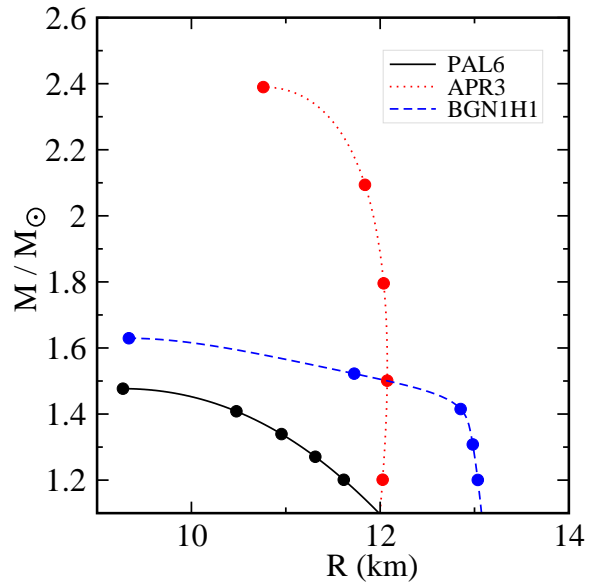}}
\caption{\label{f:mr_data_points} Exact mass-radius curves for the
  three realistic neutron-star equations of state, PAL6, APR3, and
  BGN1H1.  Large dots illustrate the data points for the spectral
  method tests that use $N_\mathrm{stars}=5$ stellar models.}
\end{figure}

We construct sets of mock $[M_i,R_i]$ data points by solving the standard
stellar structure problem using each of the exact realistic equations
of state described above.  Figure~\ref{f:mr_data_points} illustrates
these exact mass-radius data for three of the realistic equations of
state: PAL6, APR3, and BGN1H1.  We select subsets of these points for
each of our tests of the spectral inversion method.  We limit the
points used for our tests to small numbers of models (since we do not
anticipate that large numbers of observations are likely to be
available any time soon) that fall within the astrophysically relevant
range of masses: $1.2 M_\odot \leq M \leq M_\mathrm{max}$, where
$M_\mathrm{max}$ is the maximum mass star allowed for a particular
equation of state.  We choose models for our tests that are
(approximately) evenly spaced in mass within this range:
\begin{eqnarray}
M_i\approx 1.2 M_\odot  \frac{N_\mathrm{stars}-i}{N_\mathrm{stars}-1} 
+ M_\mathrm{max}\frac{i-1}{N_\mathrm{stars}-1},
\end{eqnarray}
for $i=1,...,N_\mathrm{stars}$.  The large dots on each curve in
Fig.~\ref{f:mr_data_points} illustrate the points in the data sets
with $N_\mathrm{stars}=5$ from three of these equations of state.

We have used these spectral methods to find solutions to the inverse
stellar structure problem with mass-radius data sets $[M_i,R_i]$
having $N_\mathrm{stars}=2$, 3, 4 and 5 stellar models.  In each case,
we have constructed approximate equations of state with $N_{\gamma_k}$
non-vanishing spectral parameters, for $N_{\gamma_k}=2$, ...,
$N_\mathrm{stars}$.  We use the Levenberg-Marquardt algorithm for
solving the non-linear least squares problem,
Eqs.~(\ref{e:MinimizeChi2hc}) and (\ref{e:MinimizeChi2gammak}), as
described in Ref.~\cite{numrec_f}.  We find that this method is
extremely efficient at finding solutions that minimize the
$\chi^2(h_c^i,\gamma_k)$ defined in Eq.~(\ref{e:Chi2Def}), given a
reasonably accurate initial guess for the values of the parameters
$h_c^i$ and $\gamma_k$.  Our purpose here is to explore the overall
accuracy of the spectral inversion method, and is not (at this point)
directly aimed at producing an optimal robust method for analyzing
real neutron-star observational data.  Therefore we use our knowledge
of the exact equation of state to provide the initial guesses for
$h_c^i$ and $\gamma_k$ in the test solutions reported here.  In particular
we use the values of $h_c$ for the stellar models computed from
the exact equation of state as initial guesses for $h_c^i$. Similarly
we use the values of $\gamma_k$ obtained by directly fitting
the exact equation of state data as the initial guesses for these
quantities.  While these initial guesses are not precisely the
solutions to the non-linear least squares problem, 
Eqs.~(\ref{e:MinimizeChi2hc}) and (\ref{e:MinimizeChi2gammak}), 
they are close enough that the Levenberg-Marquardt algorithm 
easily converges.

We assess the accuracy of our solutions by evaluating the differences
between the approximate equation of state $\epsilon(h,\gamma_k)$
produced by the inversion process, and the exact equation of state
$\epsilon(h)$ that was used to construct the $[M_i,R_i]$ data.
We measure these differences by constructing the error measure,
\begin{eqnarray}
\Delta^2_{N_{\gamma_k}}&=& \frac{1}{N_\mathrm{eos}}
\sum_{i=1}^{N_\mathrm{eos}}
\left[\log\left(\frac{\epsilon(h_i,\gamma_k)}{\epsilon_i}\right)\right]^2.
\label{e:DeltaDef}
\end{eqnarray}
The sum in Eq.~(\ref{e:DeltaDef}) is taken over the points,
$[\epsilon_i,h_i]$, from the tabulated realistic equation of state.  Only
the $N_\mathrm{eos}$ points that lie within the range of densities
$\epsilon_0\leq\epsilon_i\leq\epsilon_\mathrm{max}$ present in the
neutron-star models associated with a particular equation of state are
used in this sum.  Table~\ref{t:TableI} lists the values of
$\Delta_{N_{\gamma_k}}$ for each of the 34 realistic equations of
state used in our tests.  The results reported in Table~\ref{t:TableI}
are for tests that fit the maximum number of spectral parameters,
$N_{\gamma_k}=N_\mathrm{stars}$, for each set of $[M_i,R_i]$ data
points.  Fitting $N_\mathrm{stars}=2$ data points gives equations of
state approximations with average errors of only a few percent,
$\Delta_2(\mathrm{Average})=0.040$.  Using larger numbers of
$[M_i,R_i]$ data points (generally) results in higher accuracy
approximations to the equation of state, with average values of
$\Delta_3=0.029$, $\Delta_4=0.023$ and $\Delta_5=0.017$.  Thus, the
spectral approach to the inverse stellar structure problem is capable
of giving high accuracy measurements of the high density equation of
state using only a very small number of $[M_i,R_i]$ data points.
\begin{table*}[!htb]
\begin{center}
\caption{Accuracies of the neutron-star equations of state obtained by
  solving the inverse stellar structure problem.  $\Delta_N$ measures
  the average fractional error of the equation of state obtained by
  fitting $N$ different $[M_i,R_i]$ data pairs.  The parameter
  $\Upsilon_N$ measures the ratio of $\Delta_N$ to the accuracy of the
  optimal $N$-parameter spectral fit to each equation of state.  The
  parameter $\chi_N$ measures the accuracy with which the model masses
  $M$ and radii $R$ produced by the best fit equation of state match
  the exact data $M_i$ and $R_i$.
\label{t:TableI}}
\medskip
\begin{tabular}{|l|cccc|cccc|cccc|}
\hline\hline
    EOS   &$\Delta_{2}$ &$\Delta_{3}$ &$\Delta_{4}$ 
     &$\Delta_{5}$ &$\Upsilon_2$ &$\Upsilon_3$ &$\Upsilon_4$ &$\Upsilon_5$ 
     &$\chi_2$ &$\chi_3$ &$\chi_4$ &$\chi_5$ \\
\hline
    PAL6 &  0.0034 &  0.0018 &  0.0007 &  0.0003 &  1.06 &  1.09 &  1.33 &  1.91 & $ 1.1\times 10^{-9}$ & $ 7.3\times 10^{-10}$ & $ 2.0\times 10^{-9}$ & $ 1.9\times 10^{-9}$ \\
     SLY &  0.0107 &  0.0040 &  0.0022 &  0.0011 &  1.17 &  1.13 &  1.30 &  1.68 & $ 1.5\times 10^{-9}$ & $ 8.5\times 10^{-10}$ & $ 3.1\times 10^{-9}$ & $ 3.9\times 10^{-9}$ \\
     APR1 &  0.0745 &  0.0420 &  0.0225 &  0.0121 &  1.05 &  1.26 &  1.21 &  1.48 & $ 4.3\times 10^{-7}$ & $ 1.1\times 10^{-6}$ & $ 2.4\times 10^{-4}$ & $ 3.5\times 10^{-5}$ \\
     APR2 &  0.0312 &  0.0164 &  0.0093 &  0.0056 &  1.01 &  1.17 &  1.47 &  1.65 & $ 1.0\times 10^{-6}$ & $ 3.6\times 10^{-7}$ & $ 4.5\times 10^{-7}$ & $ 5.0\times 10^{-6}$ \\
     APR3 &  0.0266 &  0.0060 &  0.0030 &  0.0022 &  1.06 &  1.11 &  1.23 &  1.47 & $ 9.8\times 10^{-8}$ & $ 7.0\times 10^{-7}$ & $ 9.2\times 10^{-7}$ & $ 6.4\times 10^{-7}$ \\
     APR4 &  0.0257 &  0.0036 &  0.0017 &  0.0017 &  1.03 &  1.20 &  1.28 &  1.24 & $ 7.0\times 10^{-7}$ & $ 3.8\times 10^{-7}$ & $ 7.7\times 10^{-7}$ & $ 2.9\times 10^{-6}$ \\
     FPS &  0.0048 &  0.0061 &  0.0096 &  0.0048 &  1.06 &  1.45 &  2.52 &  2.67 & $ 1.3\times 10^{-9}$ & $ 7.7\times 10^{-9}$ & $ 4.3\times 10^{-9}$ & $ 4.7\times 10^{-9}$ \\
    WFF1 &  0.0551 &  0.0168 &  0.0220 &  0.0157 &  1.04 &  1.57 &  3.19 &  2.40 & $ 4.2\times 10^{-7}$ & $ 2.5\times 10^{-7}$ & $ 7.9\times 10^{-7}$ & $ 3.0\times 10^{-7}$ \\
    WFF2 &  0.0276 &  0.0145 &  0.0084 &  0.0055 &  1.01 &  1.21 &  1.18 &  1.45 & $ 3.2\times 10^{-7}$ & $ 2.9\times 10^{-7}$ & $ 5.5\times 10^{-7}$ & $ 8.5\times 10^{-7}$ \\
    WFF3 &  0.0126 &  0.0147 &  0.0124 &  0.0085 &  1.13 &  1.43 &  2.08 &  1.54 & $ 3.2\times 10^{-7}$ & $ 4.4\times 10^{-7}$ & $ 4.9\times 10^{-7}$ & $ 3.9\times 10^{-5}$ \\
    BBB2 &  0.0332 &  0.0328 &  0.0303 &  0.0116 &  1.01 &  1.14 &  1.39 &  1.26 & $ 4.9\times 10^{-10}$ & $ 2.3\times 10^{-9}$ & $ 7.0\times 10^{-9}$ & $ 3.5\times 10^{-5}$ \\
  BPAL12 &  0.0181 &  0.0107 &  0.0068 &  0.0032 &  1.06 &  1.08 &  1.37 &  1.43 & $ 3.2\times 10^{-9}$ & $ 3.1\times 10^{-9}$ & $ 1.9\times 10^{-9}$ & $ 1.8\times 10^{-5}$ \\
     ENG &  0.0204 &  0.0247 &  0.0200 &  0.0346 &  1.01 &  1.33 &  1.36 &  3.08 & $ 6.2\times 10^{-7}$ & $ 5.1\times 10^{-7}$ & $ 8.1\times 10^{-7}$ & $ 1.3\times 10^{-4}$ \\
    MPA1 &  0.0328 &  0.0040 &  0.0049 &  0.0049 &  1.27 &  1.23 &  1.60 &  2.15 & $ 2.6\times 10^{-7}$ & $ 7.6\times 10^{-7}$ & $ 4.4\times 10^{-7}$ & $ 2.2\times 10^{-5}$ \\
     MS1 &  0.0475 &  0.0159 &  0.0132 &  0.0008 &  1.65 &  2.79 &  3.64 &  2.21 & $ 2.6\times 10^{-6}$ & $ 2.5\times 10^{-6}$ & $ 1.8\times 10^{-6}$ & $ 3.6\times 10^{-6}$ \\
     MS2 &  0.0156 &  0.0042 &  0.0005 &  0.0005 &  1.67 &  2.00 &  2.15 &  5.98 & $ 1.5\times 10^{-10}$ & $ 2.6\times 10^{-9}$ & $ 4.6\times 10^{-10}$ & $ 1.6\times 10^{-9}$ \\
    MS1b &  0.0305 &  0.0149 &  0.0084 &  0.0017 &  1.53 &  2.33 &  2.82 &  6.20 & $ 3.5\times 10^{-7}$ & $ 7.8\times 10^{-7}$ & $ 1.8\times 10^{-6}$ & $ 1.3\times 10^{-6}$ \\
      PS &  0.1045 &  0.0789 &  0.0894 &  0.0246 &  1.66 &  2.62 &  2.97 &  1.47 & $ 6.1\times 10^{-6}$ & $ 5.2\times 10^{-6}$ & $ 1.2\times 10^{-3}$ & $ 1.7\times 10^{-4}$ \\
     GS1 &  0.0966 &  0.0586 &  0.0416 &  0.0709 &  1.08 &  1.52 &  1.10 &  2.83 & $ 9.7\times 10^{-10}$ & $ 3.1\times 10^{-10}$ & $ 2.8\times 10^{-4}$ & $ 1.7\times 10^{-4}$ \\
     GS2 &  0.0885 &  0.0911 &  0.0977 &  0.0495 &  1.46 &  2.08 &  2.25 &  1.57 & $ 3.0\times 10^{-9}$ & $ 4.7\times 10^{-9}$ & $ 1.6\times 10^{-3}$ & $ 3.5\times 10^{-4}$ \\
  BGN1H1 &  0.1352 &  0.1714 &  0.0948 &  0.1081 &  1.54 &  3.42 &  2.14 &  3.08 & $ 5.8\times 10^{-9}$ & $ 5.9\times 10^{-9}$ & $ 2.2\times 10^{-3}$ & $ 5.9\times 10^{-4}$ \\
    GNH3 &  0.0174 &  0.0186 &  0.0394 &  0.0169 &  1.27 &  1.96 &  4.78 &  2.91 & $ 2.2\times 10^{-9}$ & $ 1.9\times 10^{-9}$ & $ 4.2\times 10^{-9}$ & $ 6.0\times 10^{-9}$ \\
      H1 &  0.0294 &  0.0162 &  0.0128 &  0.0091 &  1.44 &  1.30 &  1.48 &  1.25 & $ 1.0\times 10^{-6}$ & $ 1.3\times 10^{-6}$ & $ 2.7\times 10^{-6}$ & $ 1.4\times 10^{-5}$ \\
      H2 &  0.0210 &  0.0278 &  0.0145 &  0.0091 &  1.18 &  2.00 &  2.10 &  1.32 & $ 1.0\times 10^{-6}$ & $ 1.3\times 10^{-6}$ & $ 2.8\times 10^{-6}$ & $ 1.2\times 10^{-4}$ \\
      H3 &  0.0139 &  0.0202 &  0.0177 &  0.0088 &  1.09 &  1.80 &  2.09 &  1.26 & $ 2.9\times 10^{-6}$ & $ 1.1\times 10^{-6}$ & $ 4.4\times 10^{-6}$ & $ 4.3\times 10^{-5}$ \\
      H4 &  0.0136 &  0.0251 &  0.0179 &  0.0144 &  1.32 &  2.51 &  2.69 &  2.18 & $ 5.4\times 10^{-9}$ & $ 4.8\times 10^{-9}$ & $ 4.8\times 10^{-9}$ & $ 7.7\times 10^{-5}$ \\
      H5 &  0.0139 &  0.0295 &  0.0117 &  0.0099 &  1.02 &  2.21 &  1.98 &  2.02 & $ 1.7\times 10^{-9}$ & $ 5.0\times 10^{-9}$ & $ 3.2\times 10^{-9}$ & $ 5.6\times 10^{-5}$ \\
      H6 &  0.0149 &  0.0141 &  0.0204 &  0.0158 &  1.08 &  1.03 &  1.56 &  1.39 & $ 3.3\times 10^{-9}$ & $ 3.8\times 10^{-9}$ & $ 8.8\times 10^{-9}$ & $ 8.5\times 10^{-9}$ \\
      H7 &  0.0133 &  0.0211 &  0.0123 &  0.0110 &  1.09 &  1.88 &  2.15 &  1.93 & $ 2.0\times 10^{-9}$ & $ 1.7\times 10^{-9}$ & $ 2.6\times 10^{-9}$ & $ 9.7\times 10^{-5}$ \\
    PCL2 &  0.0372 &  0.0152 &  0.0100 &  0.0100 &  1.38 &  1.14 &  1.15 &  1.28 & $ 3.8\times 10^{-7}$ & $ 1.8\times 10^{-6}$ & $ 1.7\times 10^{-6}$ & $ 1.5\times 10^{-4}$ \\
    ALF1 &  0.0796 &  0.0664 &  0.0456 &  0.0331 &  1.08 &  1.39 &  1.13 &  1.17 & $ 2.1\times 10^{-9}$ & $ 2.2\times 10^{-9}$ & $ 7.2\times 10^{-4}$ & $ 1.1\times 10^{-4}$ \\
    ALF2 &  0.0724 &  0.0600 &  0.0488 &  0.0213 &  1.04 &  1.22 &  1.76 &  1.19 & $ 4.3\times 10^{-9}$ & $ 4.5\times 10^{-9}$ & $ 6.5\times 10^{-9}$ & $ 7.8\times 10^{-5}$ \\
    ALF3 &  0.0405 &  0.0178 &  0.0185 &  0.0176 &  1.04 &  1.19 &  1.31 &  1.31 & $ 2.1\times 10^{-9}$ & $ 2.6\times 10^{-9}$ & $ 5.8\times 10^{-5}$ & $ 1.0\times 10^{-4}$ \\
    ALF4 &  0.0839 &  0.0182 &  0.0218 &  0.0171 &  1.18 &  1.35 &  2.19 &  1.81 & $ 1.6\times 10^{-8}$ & $ 7.1\times 10^{-9}$ & $ 5.8\times 10^{-9}$ & $ 1.1\times 10^{-4}$ \\
\hline
Averages &  0.0396 &  0.0289 &  0.0233 &  0.0165 &  1.22 &  1.65 &  1.97 &  2.08 & &&& \\

\hline\hline
 \end{tabular}
\end{center}
\end{table*}

Table~\ref{t:TableI} also contains two additional measures of the accuracy
of our test solutions to the inverse stellar structure problem.
One of these,
\begin{eqnarray}
\Upsilon_{N}=\frac{\Delta_{N}^{MR}}
{\Delta_{N}^{EOS}},
\label{e:UpsilonDef}
\end{eqnarray}
provides another way to measure the error in the approximate spectral
equation of state.  The quantity $\Delta_N^{MR}$ in
Eq.~(\ref{e:UpsilonDef}) refers to the error in the approximate
equation of state obtained by fitting $[M_i,R_i]$ data, as defined in
Eq.~(\ref{e:DeltaDef}).  The quantity $\Delta_N^{EOS}$ in
Eq.~(\ref{e:UpsilonDef}) refers to the error of the best possible
$N$-parameter spectral fit to this particular equation of state.  The
values of $\Delta_N^{EOS}$ for the equations of state studied here
were determined in Ref.~\cite{Lindblom10}, and are given in Table~II
of that reference.  The quantity $\Upsilon_N$ therefore measures the
accuracy of the approximate spectral equation of state obtained by
solving the inverse stellar structure problem, relative to the
accuracy of the best possible approximate $N$-parameter spectral
equation of state.  Table~\ref{t:TableI} shows that (almost) all of
these $\Upsilon_N$ measures are of order unity: the approximate
equation of state obtained with this spectral inversion method is
almost as accurate as the best possible $N$-parameter spectral fit to
the equation of state.  Table~\ref{t:TableI} also contains the values
of the quantity $\chi(h_c^i,\gamma_k)$, defined in
Eq.~(\ref{e:Chi2Def}), for each of the test solutions found here.
These values of $\chi(h_c^i,\gamma_k)$ are all much less than unity,
which shows that the least squares method is doing a good job of
minimizing the differences between the model values of
$[M(h_c^i,\gamma_k),R(h_c^i,\gamma_k)]$ and the exact data points
$[M_i,R_i]$.

\begin{figure}[htbp!]
\centerline{\includegraphics[width=3in]{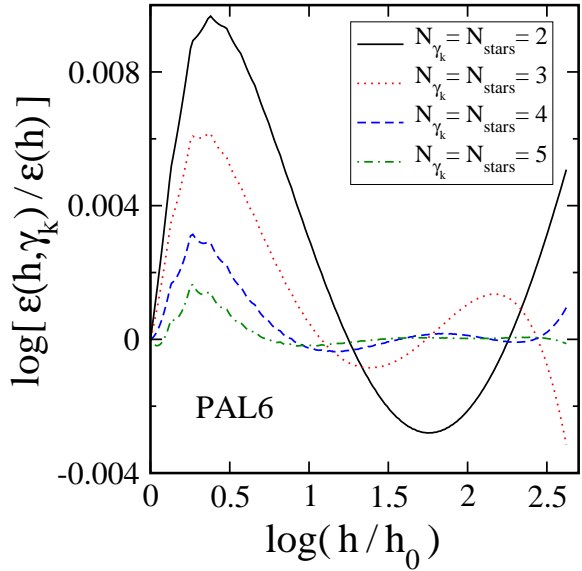}}
\caption{\label{f:pal6_comp_eos} Ratios between various approximate
  equations of state, $\epsilon(h,\gamma_k)$, obtained by fitting
  $[M,R]$ data, and the exact PAL6 equation of state, $\epsilon(h)$.
Note that $\log[\epsilon(h,\gamma_k)/\epsilon(h)]\approx
[\epsilon(h,\gamma_k)-\epsilon(h)]/\epsilon(h)$ measures the
fractional error.}
\end{figure}

\begin{figure}[t]
\centerline{\includegraphics[width=3in]{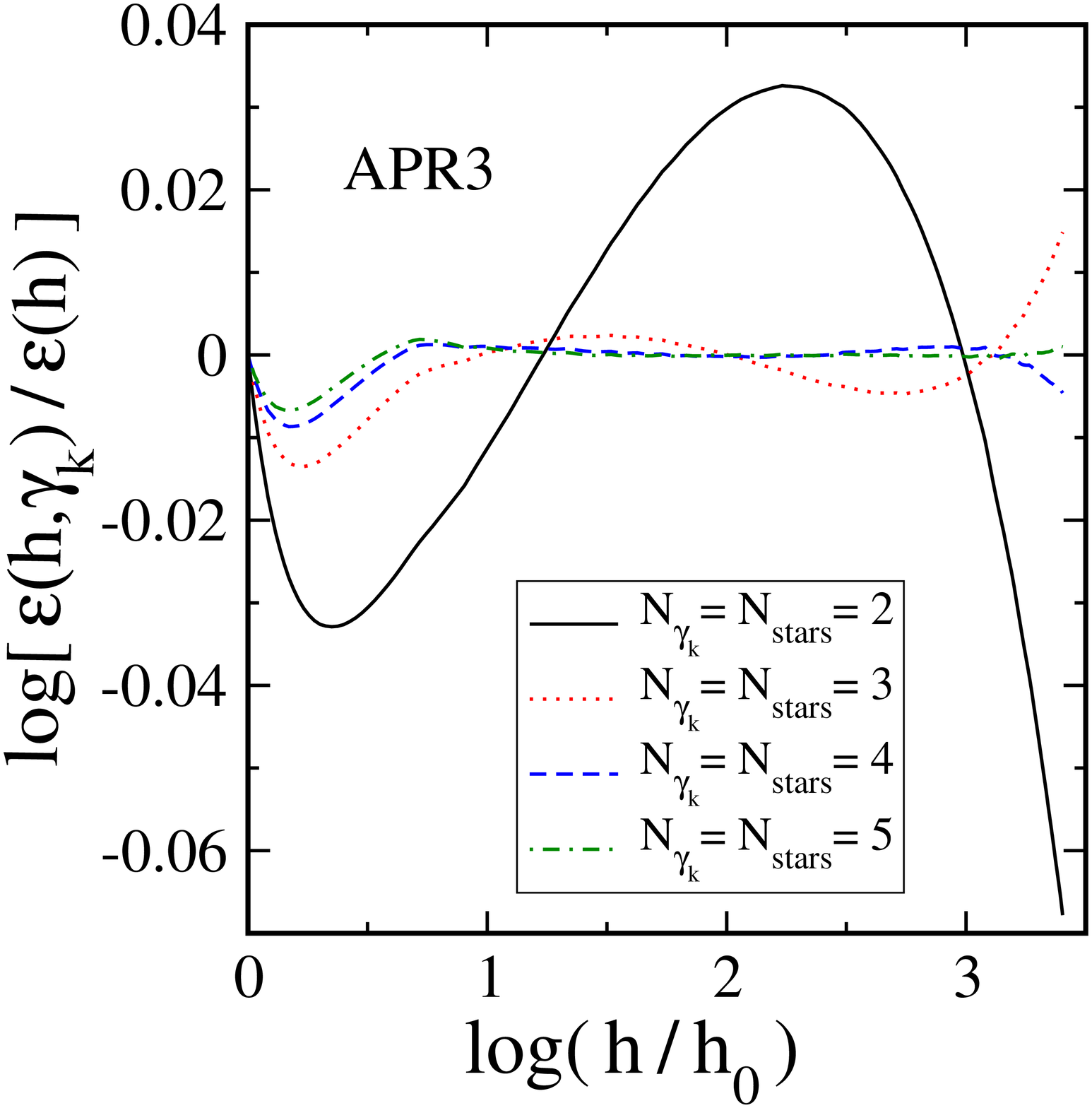}}
\caption{\label{f:apr3_comp_eos} Ratios between various approximate
  equations of state, $\epsilon(h,\gamma_k)$, obtained by fitting
  $[M,R]$ data, and the exact APR3 equation of state, $\epsilon(h)$. }
\end{figure}

\begin{figure}[t]
\centerline{\includegraphics[width=3in]{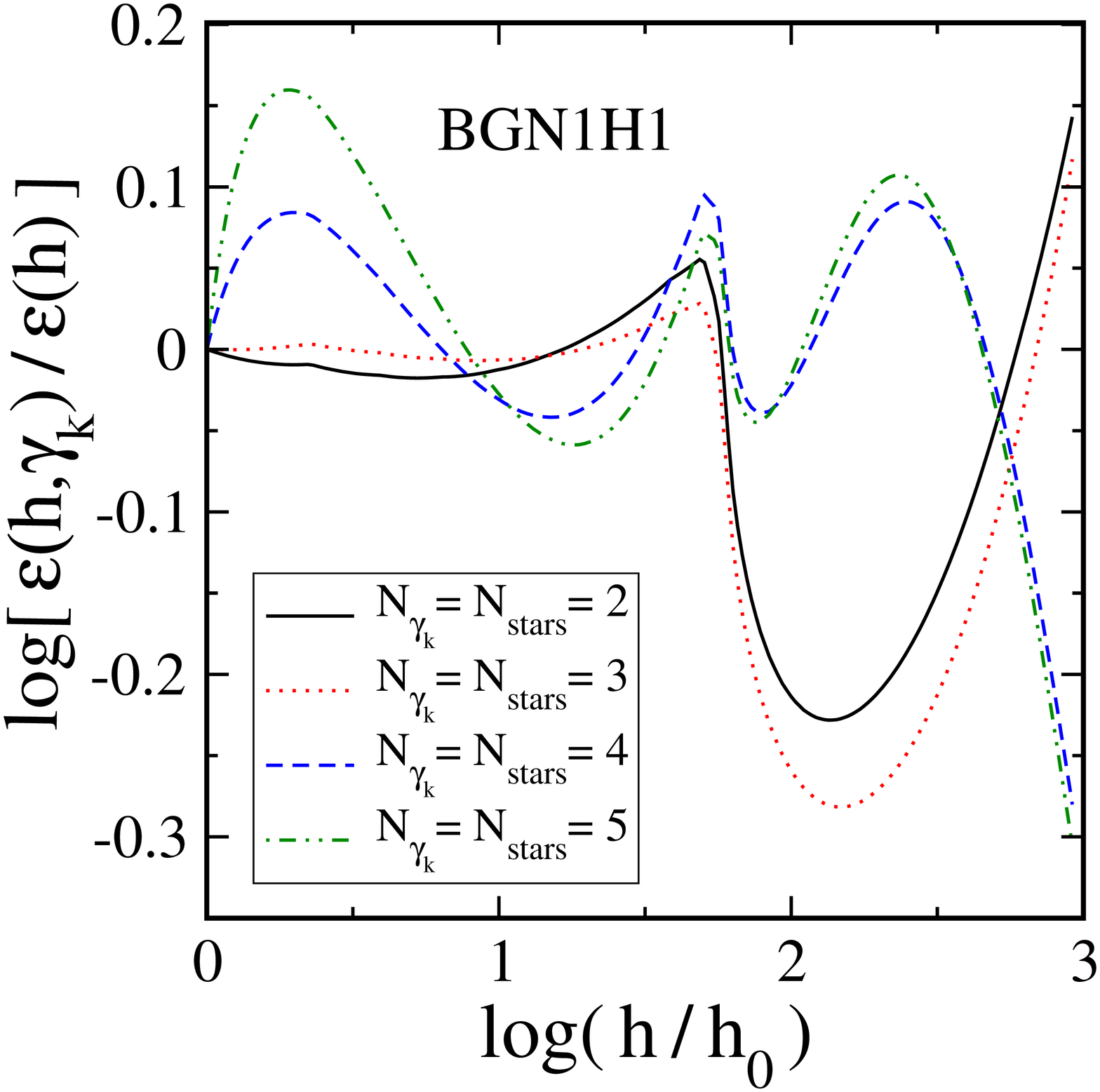}}
\caption{\label{f:bgn1h1_comp_eos} Ratios between various approximate
  equations of state, $\epsilon(h,\gamma_k)$, obtained by fitting
  $[M,R]$ data, and the exact BGN1N1 equation of state, $\epsilon(h)$.}
\end{figure}

In addition to the accuracy measures given in Table~\ref{t:TableI}, we
have made in depth studies of the errors associated with a few of
these solutions to the inverse stellar structure problem.  We have
selected for closer examination the equation of state whose
$N_\mathrm{stars}=2$ solution has the smallest error, PAL6 with
$\Delta_2=0.0034$, the equation of state whose error is the median of
the cases studied in these tests, APR3 with $\Delta_2=0.0266$, and the
equation of state having the largest error, BGN1H1 with
$\Delta_2=0.1352$.  Figure~\ref{f:pal6_comp_eos} shows the quantity
$\log[\epsilon(h,\gamma_k)/\epsilon(h)]$, which measures the
fractional difference between the best fit model equation of state
$\epsilon(h,\gamma_k)$ and the exact PAL6 equation of state
$\epsilon(h)$.  This figure shows that the errors in these solutions
to the inverse stellar structure problem become smaller as the number
of data points used in the fits with $N_{\gamma_k}=N_\mathrm{stars}$
becomes larger.  This figure also shows that the model equations of
state $\epsilon(h,\gamma_k)$ do a very good job of approximating the
actual equation of state $\epsilon(h)$ over the entire range of
densities that are present in the interiors of neutron stars.
Figures~\ref{f:apr3_comp_eos} and \ref{f:bgn1h1_comp_eos} illustrate
the analogous error measures for the equations of state obtained from
$[M_i,R_i]$ data based on the APR3 and the BGN1H1 equations of state.

These three cases illustrate the range of errors obtained using the
spectral inversion method: the best case, a typical average case, and
the worst case.  We point out that the worst case, BGN1H1, equation of
state has a strong phase transition within the neutron-star density
range.  Non-smooth equations of state of this type are difficult to
fit using spectral methods, and convergence in these cases is
typically a power law rather than exponential.  Many more spectral
parameters are therefore needed to approximate these cases accurately.
We point out, however, that the values of $\Delta_N$ for the BGN1H1
case are about ten percent, so even in this worst case the spectral
inversion method gives a reasonably accurate estimate of the equation
of state.  The values of the $\Upsilon_N$ for the BGN1H1 case all have
values below 3.5, which shows that while it is difficult to model an
equation of state of this type using a spectral fit, the spectral
inversion method nevertheless does provide a solution that is
comparable to the optimal $N_{\gamma_k}$-parameter spectral fit.

\begin{figure}[htbp!]
\centerline{\includegraphics[width=3in]{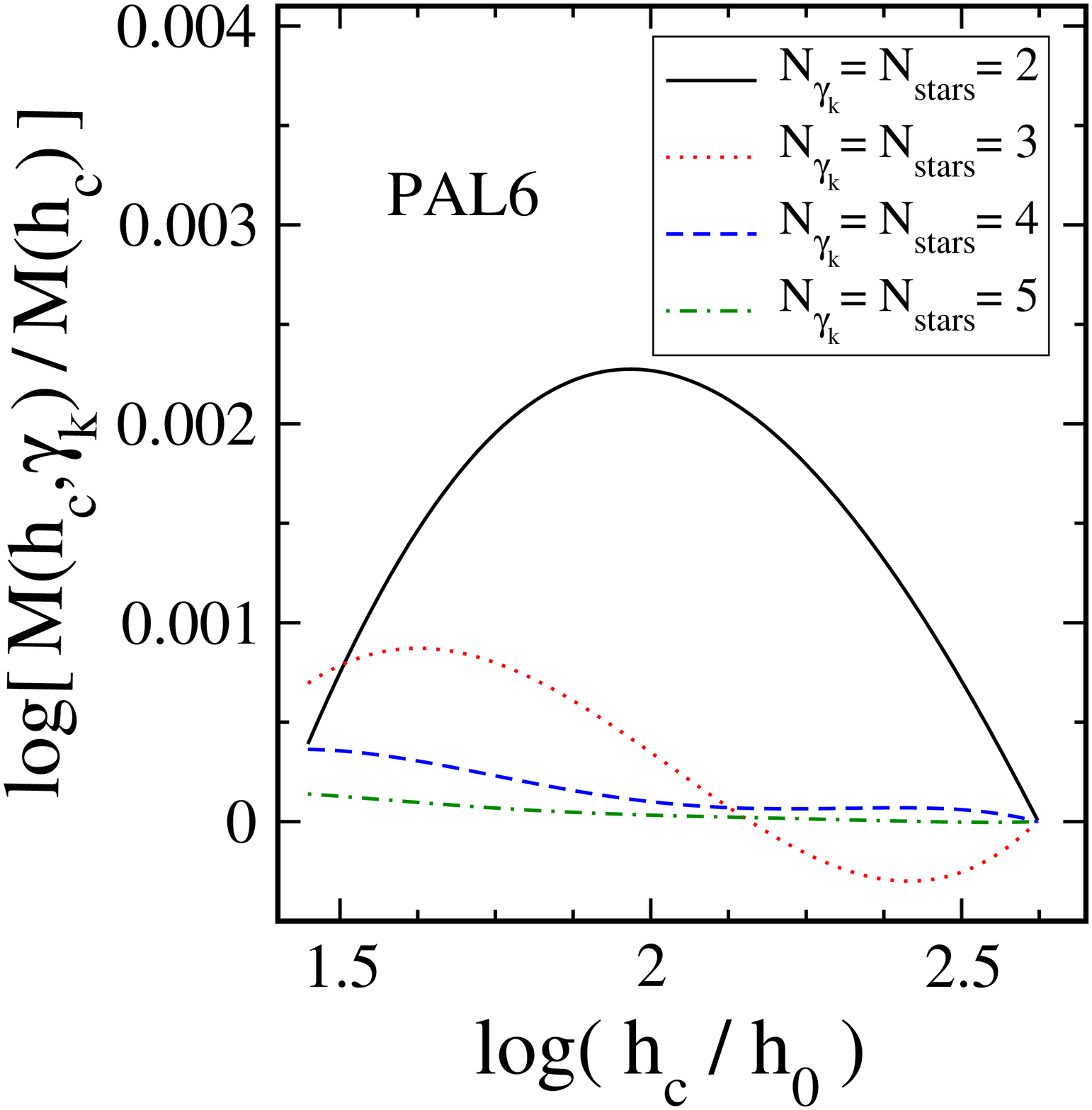}}
\caption{\label{f:pal6_comp_mr} Ratios between the masses,
  $M(h_c,\gamma_k)$ computed from the various fits, and $M(h_c)$
  computed from the exact PAL6 equation of state, for a range of
  values of the central enthalpy $h_c$ of those models.}
\end{figure}

\begin{figure}[htbp!]
\centerline{\includegraphics[width=3in]{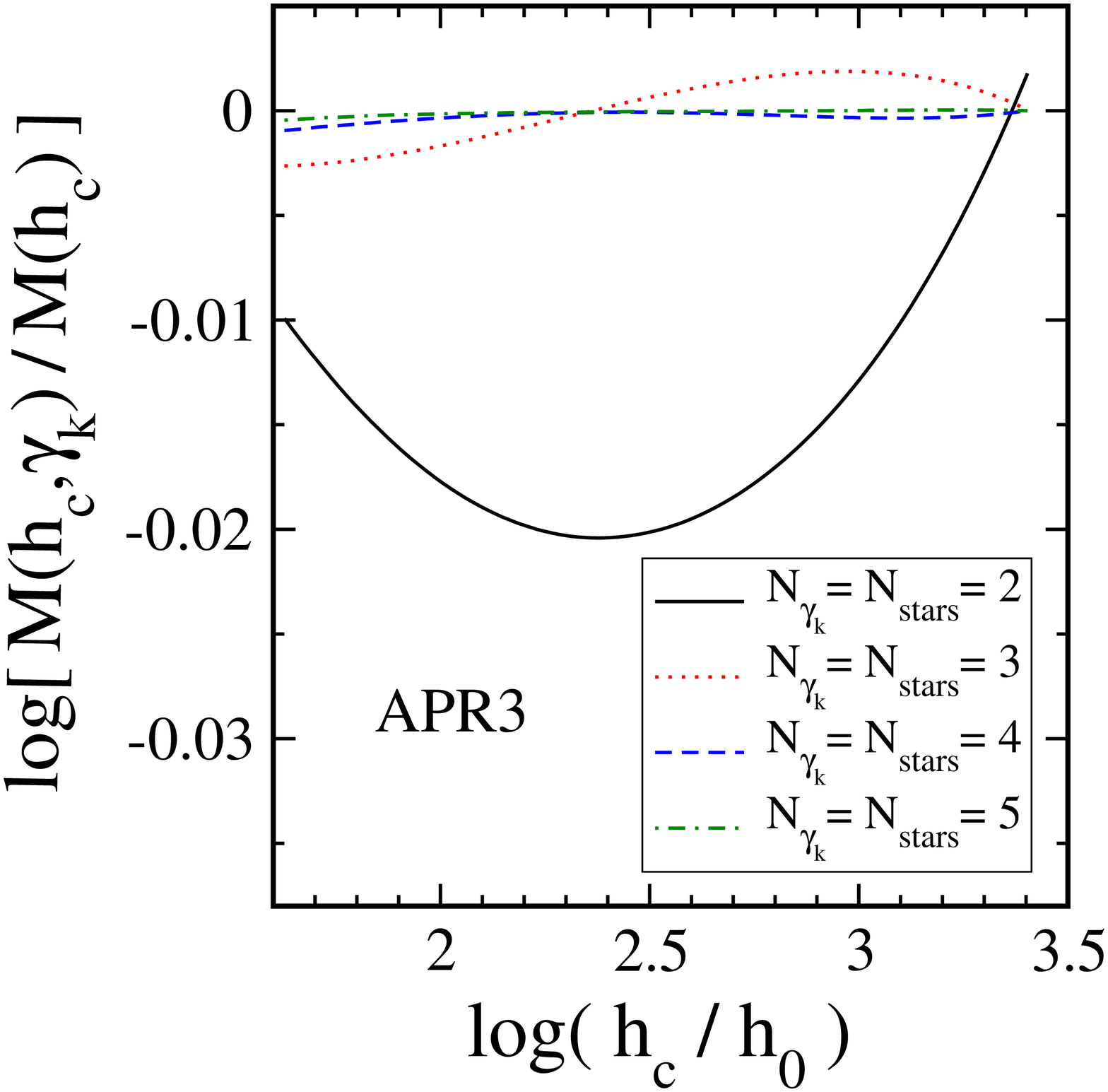}}
\caption{\label{f:apr3_comp_mr} Ratios between the masses,
  $M(h_c,\gamma_k)$ computed from the various fits, and $M(h_c)$
  computed from the exact APR3 equation of state, for a range of
  values of the central enthalpy $h_c$ of those models.}
\end{figure}

\begin{figure}[htbp!]
\centerline{\includegraphics[width=3in]{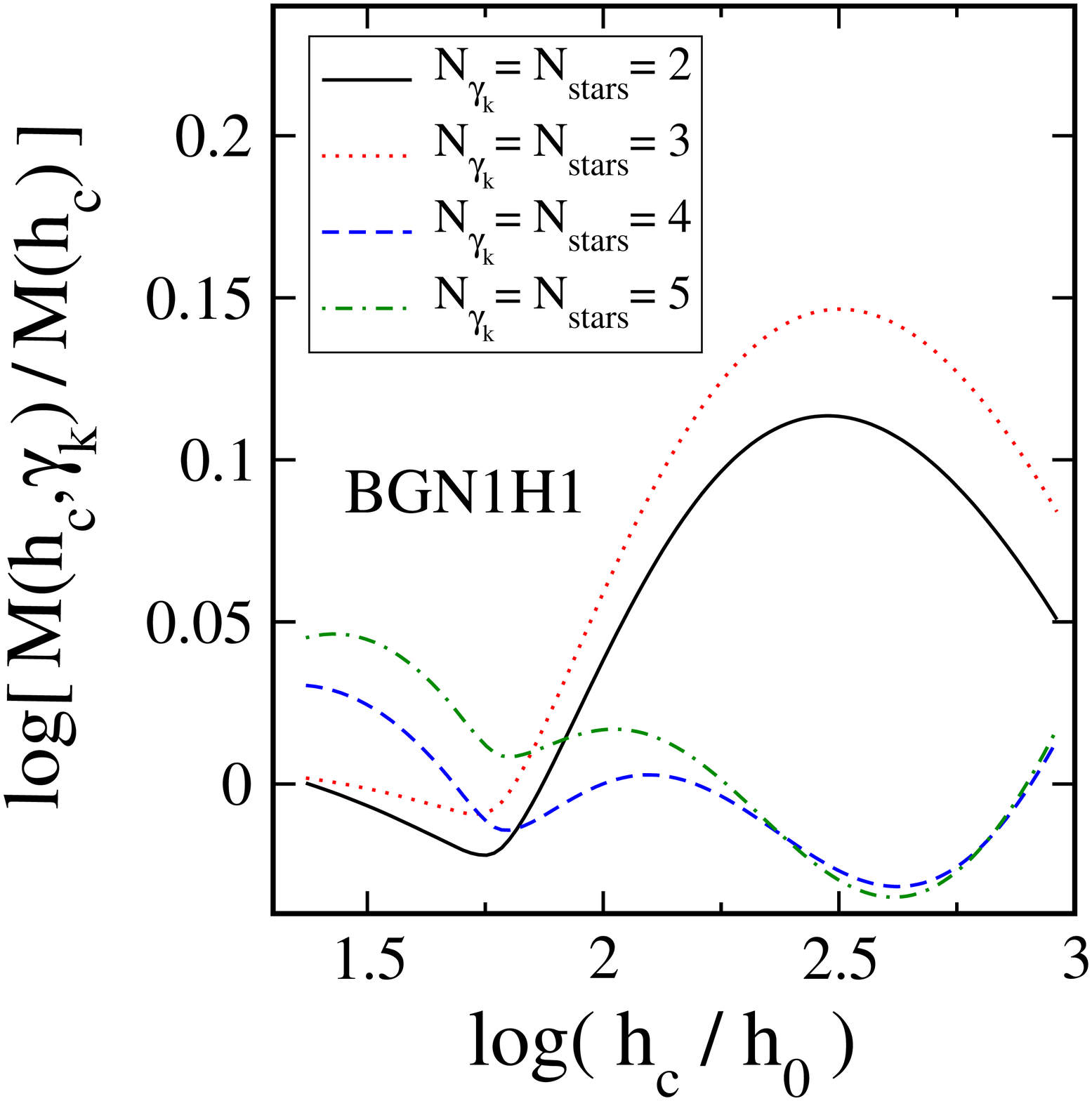}}
\caption{\label{f:bgn1h1_comp_mr} Ratios between the masses,
  $M(h_c,\gamma_k)$ computed from the various fits, and $M(h_c)$
  computed from the exact BGN1H1 equation of state, for a range of
  values of the central enthalpy $h_c$ of those models.}
\end{figure}

Given an approximate spectral equation of state,
$\epsilon(h,\gamma_k)$, we can use it to compute the complete
mass-radius curve $[M(h_c,\gamma_k),R(h_c,\gamma_k)]$ for the full
range of central enthalpies, $h_c$.  This model mass-radius curve
should agree with the exact curve $[M(h),R(h)]$ very well, at least at
the points $[M_i,R_i]$ used in the inversion process.  However, they
will not agree everywhere, and the size of the differences is another
measure of how well the approximate equation of state agrees with the
exact.  Figure~\ref{f:pal6_comp_mr} illustrates the differences
between the model masses $M(h_c,\gamma_k)$ and the exact masses
$M(h_c)$ for the PAL6 equation of state.\footnote{Note that the masses
  in Figs.~\ref{f:pal6_comp_mr}--~\ref{f:bgn1h1_comp_mr} are compared
  between models having the same central enthalpy $h_c$.  The central
  enthalpy of the exact model with $M=M_i$ need not be exactly equal
  to the central enthalpy of the best fit model, $h_c^i$.  Therefore
  these curves need not have zeros at those points.}
Figures~\ref{f:apr3_comp_mr} and \ref{f:bgn1h1_comp_mr} make similar
comparisons for the APR3 and the BGN1H1 cases.  We note that the error
measures, $\log[M(h_c,\gamma_k)/M(h_c)]$, shown in
Figs.~\ref{f:pal6_comp_mr}--\ref{f:bgn1h1_comp_mr}, are somewhat
smaller in size than the error measures,
$\log[\epsilon(h,\gamma_k)/\epsilon(h)]$, shown in
Figs.~\ref{f:pal6_comp_eos}--\ref{f:bgn1h1_comp_eos}.  These error
measures provide one more piece of evidence that the spectral
solutions to the inverse stellar structure problem are quite accurate.

\section{Discussion}
\label{s:Discussion}

In summary, we have developed a new method for solving the
relativistic inverse stellar structure problem based on the
construction of a spectral expansion of the unknown high density part
of the equation of state of the star.  The results of our numerical
tests of this new method, described in
Sec.~\ref{s:TestingSpectralInversionMethod}, are quite
impressive. Using only two $[M_i,R_i]$ data points, this new method
can determine the entire high density part of the neutron-star
equation of state with errors (on average) of just a few percent.  The
addition of more data points (generally) results in higher accuracy
approximations.  We also show that $N$-parameter spectral
approximations to the equation of state determined in this way are
almost as accurate as the best possible $N$-parameter spectral
approximations.  This is quite remarkable.  It shows that macroscopic
mass-radius measurements are strongly correlated to the properties of
the equation of state, and such measurements should therefore allow us
(eventually) to measure the high density part of the neutron-star
equation of state with great precision.

A close inspection of the results from the various tests summarized in
Table~\ref{t:TableI} reveals a number of anomalies that merit further
study.  For example, the error measure $\Delta_{N_{\gamma_k}}$,
defined in Eq.~(\ref{e:DeltaDef}), is expected to decrease as the
number of spectral parameters $N_{\gamma_k}$ is increased, i.e. that
$\Delta_{N_{\gamma_k}}\geq\Delta_{N_{\gamma_k}+1}$.  This seems to be
true for most of our tests, but there are also a number of exceptions
in Table~\ref{t:TableI}.  The equation of state FPS, for example, has
$\Delta_2=0.0048$, $\Delta_3=0.0061$, $\Delta_4=0.0096$, and
$\Delta_5=0.0048$.  What is going on?  Such a sequence of errors would
be consistent, for example, with the idea that this particular
equation of state is not well represented by these low order spectral
expansions, i.e. that these expansions in this case are not yet in the
convergent regime.  This does not seem to be the case however since
the optimal spectral fits to the FPS equation of state do appear to be
convergent with these same numbers of spectral parameters, cf. Table
II of Ref.~\cite{Lindblom10}.  Another (more likely) explanation of
the anomalous results found in Table~\ref{t:TableI} is that the minima
of $\chi^2(h_c,\gamma_k)$ found by the Levenberg-Marquardt algorithm
for these cases are just local minima and not the desired global
minima.  An interesting area for further research on this problem,
therefore, will be to explore the use of more robust numerical methods
for finding global minima of complicated non-linear functions like
$\chi^2(h_c,\gamma_k)$.

Another interesting direction for future research on this problem will
be to explore how robust this kind of solution to the inverse stellar
structure problem will be when applied to more realistic $[M_i,R_i]$
data sets.  The data used here were idealized in two important ways.
First, the mock $[M_i,R_i]$ data used here were supplied with very
high precision.  Real astrophysical measurements of these quantities
will have significant errors.  How will measurement errors influence
the accuracy of the equation of state that is constructed by these
techniques?  Second, the mock $[M_i,R_i]$ data used here were chosen
to cover uniformly the astrophysically relevant range of neutron-star
masses.  Real astrophysical measurements will not be distributed in
such an orderly way.  How will the accuracy of the implied equation of
state be affected by different, presumably less ideal, data
distributions?

The version of the inverse stellar structure problem studied here is
based on the use of mass $M$ and radius $R$ measurements to determine
the high density part of the equation of state.  These are not the
only macroscopic properties of neutron stars that could potentially be
measured.  It is not too difficult to imagine for example that the
moment of inertias or the tidal Love numbers might be more easily
observable for some types of observations.  Another interesting
direction for future study will therefore be to explore the use of
other measurement data, say the mass and Love number (which could be
measured using gravitational wave observations of neutron-star
mergers), as input for solving the inverse stellar structure problem
using the spectral methods developed here.


\acknowledgments We thank John Friedman, Benjamin Lackey, and Benjamin
Owen for helpful discussions about this work, and J.F. and B.L. for
providing tables of the various realistic equations of state.  A
portion of this research was carried out during the time L.L. was a
visitor at the Leonard E. Parker Center for Gravitation, Cosmology and
Astrophysics, University of Wisconsin at Milwaukee.  This research was
supported in part by a grant from the Sherman Fairchild Foundation, by
NSF grants PHY1005655 and DMS1065438, and by NASA grant NNX09AF97G.


\appendix
 
\section{Computing Derivatives of $M$ and $R$.}
\label{s:AppendixA}

This Appendix describes how the derivatives of the total masses
$M(h_c,\gamma_k)$ and radii $R(h_c,\gamma_k)$ are computed with
respect to the parameters $h_c$ and $\gamma_k$.  To begin, however, we
present a little more detail on how the alternate form of the stellar
structure Eqs.~(\ref{e:OVEquation_m}) and (\ref{e:OVEquation_r}) are
solved numerically.  These equations are,
\begin{eqnarray}
\frac{dm}{dh}&=&{\cal M}(m,r,\epsilon,p)\equiv
-\frac{4\pi r^3 \epsilon(r-2m)}{m+4\pi r^3 p},
\label{e:OVEquation_M}\\
\frac{dr}{dh}&=&{\cal R}(m,r,p)\equiv
-\frac{r(r-2m)}{m+4\pi r^3 p}
\label{e:OVEquation_R},
\end{eqnarray}
where the quantities ${\cal M}(m,r,\epsilon,p)$ and ${\cal R}(m,r,p)$
merely represent the expressions on the right sides.  These equations
are solved numerically by specifying conditions, $m(h_c)=r(h_c)=0$, at
the center of the star where $h=h_c$ and then integrating out to the
surface of the star where $h=0$.  Like the standard
Oppenheimer-Volkoff version of the problem, Eqs.~(\ref{e:OVm}) and
(\ref{e:OVp}), the right sides of Eqs.~(\ref{e:OVEquation_M}) and
(\ref{e:OVEquation_R}), i.e. the functions ${\cal M}(m,r,\epsilon,p)$
and ${\cal R}(m,r,p)$, have the ill behaved form $0/0$ there.
Consequently it is necessary to start any numerical integration of
these equations slightly away from the singular point $h=h_c$.  The
needed starting conditions can be obtained using a power series
solution to the equations.  The needed power series are given in
Eqs.~(7) and (8) of Ref.~\cite{Lindblom1992}, and can be written in the
form,
\begin{eqnarray}
r(h) &=& r_1 (h_c - h)^{1/2} + r_3 (h_c -h )^{3/2} 
\nonumber\\
&&\quad+ {\cal O}(h_c - h)^{5/2},\label{e:h_series_r}\\
m(h) &=& m_3 (h_c - h)^{3/2} + m_5 (h_c - h)^{5/2} 
\nonumber\\
&&\quad+ {\cal O}(h_c - h)^{7/2},\label{e:h_series_m}
\end{eqnarray}
where $r_1$, $r_3$, $m_3$ and $m_5$ are given by
\begin{eqnarray}
r_1 &=& \left[\frac{3}{2\pi(\epsilon_c + 3 p_c)}\right]^{1/2},\\
r_3 &=& - \frac{r_1}{4(\epsilon_c+3p_c)}
\left[\epsilon_c - 3p_c -\frac{3(\epsilon_c+p_c)^2}{5 p_c \Gamma_c}\right],\\
m_3 &=& \frac{4\pi}{3} \epsilon_c r_1^3,\\
m_5 &=& 4\pi r_1^3\left[\frac{r_3\epsilon_c}{r_1} -
\frac{(\epsilon_c+p_c)^2}{5 p_c\Gamma_c}\right].
\end{eqnarray}
The quantities $\epsilon_c$, $p_c$ and $\Gamma_c$ in these expressions
are the energy density, pressure and the adiabatic index evaluated at
the center of the star where $h=h_c$, $\epsilon_c=\epsilon(h_c)$,
$p_c=p(h_c)$, and $\Gamma_c =\Gamma(h_c)$.

It will be useful for our least-squares minimization problem to know
how the solutions to Eqs.~(\ref{e:OVEquation_M}) and
(\ref{e:OVEquation_R}) change as the parameters $h_c$ and $\gamma_k$
are varied.  Let $\lambda$ denote any one of the parameters:
$\lambda=\{h_c,\gamma_k\}$.  We wish to derive equations for the
derivatives of the solutions to these equations with respect to these
parameters: $\partial m/\partial\lambda$ and $\partial
r/\partial\lambda$.  It is straightforward to determine the needed
auxiliary equations by differentiating, Eqs.~(\ref{e:OVEquation_M})
and (\ref{e:OVEquation_R}) with respect to $\lambda$:
\begin{eqnarray}
\frac{d}{dh}\left(\frac{\partial m}{\partial\lambda}\right)&=&
\frac{\partial {\cal M}}{\partial m}\frac{\partial m}{\partial\lambda}
+\frac{\partial {\cal M}}{\partial r}\frac{\partial r}{\partial\lambda}
+\frac{\partial {\cal M}}{\partial \epsilon}\frac{\partial \epsilon}{\partial\lambda}\nonumber \\
&&+\frac{\partial {\cal M}}{\partial p}\frac{\partial p}{\partial\lambda},
\label{e:VarOVEquation_m}\qquad\\
\frac{d}{dh}\left(\frac{\partial r}{\partial\lambda}\right)&=&
\frac{\partial {\cal R}}{\partial m}\frac{\partial m}{\partial\lambda}
+\frac{\partial {\cal R}}{\partial r}\frac{\partial r}{\partial\lambda}
+\frac{\partial {\cal R}}{\partial p}\frac{\partial p}{\partial\lambda}.
\qquad\qquad
\label{e:VarOVEquation_r}\end{eqnarray}
The various derivatives $\partial {\cal M}/\partial m$, etc. are determined
directly from the stellar structure equations, Eqs.~(\ref{e:OVEquation_M})
and (\ref{e:OVEquation_R}):
\begin{eqnarray}
\frac{\partial {\cal M}}{\partial m}&=& \frac{8\pi r^3\epsilon - {\cal M}}{m+4\pi r^3 p},\\
\frac{\partial {\cal M}}{\partial r}&=& 
-4\pi r^2\frac{3p{\cal M}+2\epsilon(2r-3m)}{m+4\pi r^3 p},\\
\frac{\partial {\cal M}}{\partial p}&=& 
-\frac{4\pi r^3 {\cal M}}{m+4\pi r^3 p},\\
\frac{\partial {\cal M}}{\partial \epsilon}&=& 
-\frac{4\pi r^3(r-2m)}{m+4\pi r^3 p},\\
\frac{\partial {\cal R}}{\partial m}&=& \frac{2r - {\cal R}}{m+4\pi r^3 p},\\
\frac{\partial {\cal R}}{\partial r}&=& 
-\frac{12\pi r^2p{\cal R}+2(r-m)}{m+4\pi r^3 p},\\
\frac{\partial {\cal R}}{\partial p}&=& 
-\frac{4\pi r^3 {\cal R}}{m+4\pi r^3 p}.
\end{eqnarray}

For the
case when $\lambda=\gamma_k$, the derivatives
$\partial\epsilon/\partial\gamma_k$ and $\partial p/\partial\gamma_k$
are determined from Eqs.~(\ref{e:PressueH})--(\ref{e:TildeMuDef}).
The needed expressions are given by:
\begin{eqnarray}
\frac{\partial\tilde\mu(h)}{\partial\gamma_k}&=&
\int_{h_0}^h\left[\log\left(\frac{h'}{h_0}\right)\right]^k
\frac{e^{h'}dh'}{\Gamma(h')},\\
\frac{\partial p(h)}{\partial\gamma_k}&=& -p(h)
\int_{h_0}^h\frac{\partial\tilde\mu(h')}{\partial\gamma_k}
\frac{e^{h'}dh'}{\left[\tilde\mu(h')\right]^2},
\label{e:dpdgammak}\\
\frac{\partial \epsilon(h)}{\partial\gamma_k}&=& 
\frac{\partial p(h)}{\partial\gamma_k}\frac{\epsilon(h)}{p(h)}-
\frac{\partial\tilde\mu(h)}{\partial\gamma_k}
\frac{e^{h}p(h)}{\left[\tilde\mu(h)\right]^2}.
\label{e:depsilondgammak}
\end{eqnarray}
The integrals needed to determine these quantities can be performed
numerically with good efficiency and accuracy using Gaussian
quadrature.  The equation of state does not depend on the parameter
$h_c$, and so $\partial\epsilon/\partial h_c=\partial p/\partial
h_c=0$.  Consequently the equations that determine $\partial
m/\partial h_c$ and $\partial r/\partial h_c$ in
Eqs.~(\ref{e:VarOVEquation_m}) and (\ref{e:VarOVEquation_r}) are
somewhat simpler than those for $\partial m/\partial\gamma_k$ and
$\partial r/\partial\gamma_k$.

The functions $\partial m/\partial \lambda$ and $\partial r/\partial
\lambda$ are determined by solving Eqs.~(\ref{e:VarOVEquation_m}) and
(\ref{e:VarOVEquation_r}) numerically.  This can be done by
integrating them from the center of the star where $h=h_c$ out to the
surface of the star where $h=0$.  To do this we need to impose the
appropriate boundary conditions for these functions at $h=h_c$.  The
needed boundary conditions can be found by differentiating the power
series solutions, Eqs.~(\ref{e:h_series_r}) and (\ref{e:h_series_m}),
with respect to the parameters $\lambda$.  The quantities $r_1$,
$r_3$, $m_3$ and $m_5$, which appear in these power series solutions,
depend on the central values of the thermodynamic quantities
$\epsilon_c=\epsilon(h_c)$, $p_c= p(h_c)$, and $\Gamma_c=\Gamma(h_c)$,
and through them the parameters $\lambda=\{h_c,\gamma_k\}$.  For the
case where $\lambda=\gamma_k$ these derivatives can be written as
\begin{eqnarray}
&&\!\!\!\!\!
\frac{\partial r(h)}{\partial\gamma_k} =
\left[\frac{\partial r_1}{\partial \epsilon_c}
\frac{\partial\epsilon_c}{\partial\gamma_k}
+\frac{\partial r_1}{\partial p_c}
\frac{\partial p_c}{\partial\gamma_k}\right](h_c-h)^{1/2}\nonumber\\
&&\quad+\left[\frac{\partial r_3}{\partial \epsilon_c}
\frac{\partial\epsilon_c}{\partial\gamma_k}
+\frac{\partial r_3}{\partial p_c}
\frac{\partial p_c}{\partial\gamma_k}
+\frac{\partial r_3}{\partial \Gamma_c}
\frac{\partial \Gamma_c}{\partial\gamma_k}\right](h_c-h)^{3/2}\nonumber\\
&&\quad+ {\cal O}(h_c - h)^{5/2},
\label{e:h_series_dr_gammak}\\
&&\!\!\!\!\!
\frac{\partial m(h)}{\partial\gamma_k} =
\left[\frac{\partial m_3}{\partial \epsilon_c}
\frac{\partial\epsilon_c}{\partial\gamma_k}
+\frac{\partial m_3}{\partial p_c}
\frac{\partial p_c}{\partial\gamma_k}\right](h_c-h)^{3/2}\nonumber\\
&&\quad+\left[\frac{\partial m_5}{\partial \epsilon_c}
\frac{\partial\epsilon_c}{\partial\gamma_k}
+\frac{\partial m_5}{\partial p_c}
\frac{\partial p_c}{\partial\gamma_k}
+\frac{\partial m_5}{\partial \Gamma_c}
\frac{\partial \Gamma_c}{\partial\gamma_k}\right](h_c-h)^{5/2}\nonumber\\
&&\quad+ {\cal O}(h_c - h)^{7/2}.
\label{e:h_series_dm_gammak}
\end{eqnarray}
The derivatives of $r_1$, $r_3$, $m_3$ and $m_5$ with respect to the
parameters $\epsilon_c$, $p_c$ and $\Gamma_c$ which appear in
Eqs.~(\ref{e:h_series_dr_gammak}) and (\ref{e:h_series_dm_gammak}) are
given by:
\begin{eqnarray}
\frac{\partial r_1}{\partial \epsilon_c}&=&
-\frac{r_1}{2(\epsilon_c+3p_c)},
\label{e:dr1_depsilonc}\\
\frac{\partial r_1}{\partial p_c}&=&
3\frac{\partial r_1}{\partial \epsilon_c}.\\
\frac{\partial r_3}{\partial \epsilon_c}&=&
\frac{r_3}{r_1}\frac{\partial r_1}{\partial \epsilon_c}
-\frac{r_1}{4(\epsilon_c+3p_c)}
\left[1 +\frac{4r_3}{r_1}
-\frac{6(\epsilon_c+3p_c)}{5p_c\Gamma_c}\right],\nonumber\\
&&\\
\frac{\partial r_3}{\partial p_c}&=&
\frac{r_3}{r_1}\frac{\partial r_1}{\partial p_c}
+\frac{3r_1}{4(\epsilon_c+3p_c)}
\left[1 -\frac{4r_3}{r_1}
-\frac{\epsilon_c^2-p_c^2}{5p_c^2\Gamma_c}\right],\nonumber\\
&&\\
\frac{\partial r_3}{\partial \Gamma_c}&=&
-\frac{3r_1(\epsilon_c+p_c)^2}{20p_c(\epsilon_c+3p_c)\Gamma_c^2},\\
\frac{\partial m_3}{\partial \epsilon_c}&=&\frac{4\pi}{3}r_1^3\left[1+
\frac{3\epsilon_c}{r_1}\frac{\partial r_1}{\partial\epsilon_c}\right],\\
\frac{\partial m_3}{\partial p_c}&=&4\pi\epsilon_cr_1^2
\frac{\partial r_1}{\partial p_c},
\end{eqnarray}
\begin{eqnarray}
\frac{\partial m_5}{\partial \epsilon_c}&=&4\pi r_1^2\left[r_3
+\frac{2\epsilon_cr_3}{r_1}\frac{\partial r_1}{\partial\epsilon_c}
+\epsilon_c\frac{\partial r_3}{\partial\epsilon_c}\right]\nonumber\\
&&-\frac{4\pi r_1^2(\epsilon_c+p_c)}{5p_c\Gamma_c}
\left[
2r_1+3(\epsilon_c+p_c)\frac{\partial r_1}{\partial\epsilon_c}\right],
\qquad\\
\frac{\partial m_5}{\partial p_c}&=&4\pi \epsilon_c r_1^2\left[
\frac{2r_3}{r_1}\frac{\partial r_1}{\partial p_c}+
\frac{\partial r_3}{\partial p_c}\right]\nonumber\\
&&+\frac{4\pi r_1^3(\epsilon_c+p_c)}{5p_c^2\Gamma_c}
\left[
\epsilon_c -\frac{3p_c(\epsilon_c+p_c)}{r_1}
\frac{\partial r_1}{\partial p_c}\right],
\label{e:dm5_dpc}
\\
\frac{\partial m_5}{\partial \Gamma_c}&=&
4\pi r_1^3\left[\frac{\epsilon_c}{r_1}
\frac{\partial r_3}{\partial\Gamma_c}+
\frac{(\epsilon_c+p_c)^2}{5 p_c\Gamma_c^2}\right].
\label{e:dm5_dGammac}
\end{eqnarray}
The values of the derivatives $\partial p_c/\partial\gamma_k$ and
$\partial \epsilon_c/\partial\gamma_k$ are obtained by evaluating
Eqs.~(\ref{e:dpdgammak}) and (\ref{e:depsilondgammak}) at $h=h_c$,
while the derivative $\partial\Gamma_c/\partial \gamma_k$ is given
by
\begin{eqnarray}
\frac{\partial\Gamma_c}{\partial\gamma_k} = 
\left[\log\left(\frac{h_c}{h_0}\right)\right]^k\Gamma_c.
\end{eqnarray}

For the case where $\lambda=h_c$ the expressions for the derivatives
$\partial r/\partial \lambda$ and $\partial m/\partial \lambda$ have
somewhat different forms because $h_c$ appears explicitly in
expansions in Eqs.~(\ref{e:h_series_r}) and (\ref{e:h_series_m}).
Differentiating these series with respect to $h_c$, keeping only
the two leading terms, gives
\begin{eqnarray}
&&\frac{\partial r(h)}{\partial h_c} =
\frac{r_1}{2}(h_c-h)^{-1/2}\nonumber\\
&&\qquad+
\left[\frac{\partial r_1}{\partial \epsilon_c}
\frac{\partial\epsilon_c}{\partial h_c}
+\frac{\partial r_1}{\partial p_c}
\frac{\partial p_c}{\partial h_c}+\frac{3 r_3}{2}
\right](h_c-h)^{1/2}\nonumber\\
&&\qquad+ {\cal O}(h_c - h)^{3/2},\label{e:h_series_dr_hc}\\
&&\frac{\partial m(h)}{\partial h_c} =
\frac{3 m_3}{2}(h_c-h)^{1/2}\nonumber\\
&&\qquad+\left[\frac{\partial m_3}{\partial \epsilon_c}
\frac{\partial\epsilon_c}{\partial h_c}
+\frac{\partial m_3}{\partial p_c}
\frac{\partial p_c}{\partial h_c}
+\frac{5 m_5}{2}\right](h_c-h)^{3/2}\nonumber\\
&&\qquad+ {\cal O}(h_c - h)^{5/2}.
\label{e:h_series_dm_hc}
\end{eqnarray}
The derivatives of $r_1$, $r_3$, $m_3$ and $m_5$ with respect to the
parameters $\epsilon_c$ and $p_c$ which appear in
Eqs.~(\ref{e:h_series_dr_hc}) and (\ref{e:h_series_dm_hc}) are given
as before by the expressions in
Eqs.~(\ref{e:dr1_depsilonc})--(\ref{e:dm5_dpc}).  The derivatives
$\partial\epsilon_c/\partial h_c$ and $\partial p_c/\partial h_c$
which appear in Eqs.~(\ref{e:h_series_dr_hc}) and
(\ref{e:h_series_dm_hc}) are obtained
by evaluating Eqs.~(\ref{e:dpdh}) and (\ref{e:depsilondh})
at $h=h_c$:
\begin{eqnarray}
\frac{\partial p_c}{\partial h_c} &=& \epsilon_c + p_c,\label{e:dpcdhc}\\
\frac{\partial\epsilon_c}{\partial h_c} &=& \frac{(\epsilon_c + p_c)^2}{p_c\Gamma(h_c)}.
\label{e:depsiloncdhc}
\end{eqnarray}

\section{Interpolating the Exact Equation of State}
\label{s:AppendixB}

We are often presented with an ``exact'' equation of state that is
represented as a table of energy densities $\epsilon_i$ and the
corresponding pressures $p_i$.  For our purposes here we will convert
these to an equation of state of the form $\epsilon=\epsilon(h)$ and
$p=p(h)$ in the following way.  We do this by assuming that the exact
equation of state is obtained for values intermediate between those
given in the table, $\epsilon_i \leq \epsilon\leq \epsilon_{i+1}$, 
by the interpolation formula:
\begin{eqnarray}
\frac{p}{p_i} &=& \left(\frac{\epsilon}{\epsilon_i}\right)^{c_{i+1}},
\label{e:piInterpolate}\\
c_{i+1}&=&\frac{\log(p_{i+1}/p_i)}{\log(\epsilon_{i+1}/\epsilon_i)}.
\end{eqnarray}
For smaller values of the density, $\epsilon\leq \epsilon_1$, we
assume:
\begin{eqnarray}
\frac{p}{p_1} &=& \left(\frac{\epsilon}{\epsilon_1}\right)^{c_1},
\label{e:p1Interpolate}\\
c_1&=&\frac{\log(p_{2}/p_1)}{\log(\epsilon_{2}/\epsilon_1)}.
\end{eqnarray}

Given this prescription for interpolation, it is straightforward to
show that the values of the enthalpy
\begin{eqnarray}
h(p)=\int_0^p\frac{dp'}{\epsilon(p')+p'},
\label{e:AppendixEnthalpy}
\end{eqnarray}
are given at the table entry values $h_i=h(p_i)$, by
\begin{eqnarray}
h_1&=& \frac{c_1}{c_1-1}\log\left(\frac{\epsilon_1+p_1}{\epsilon_1}\right),\\
h_{i+1}&=& h_i+\frac{c_{i+1}}{c_{i+1}-1}
\log\left[\frac{\epsilon_i(\epsilon_{i+1}+p_{i+1})}
{\epsilon_{i+1}(\epsilon_i+p_i)}\right].
\end{eqnarray}

The pressure is determined as a function of the enthalpy, by
performing the integral in Eq.~(\ref{e:AppendixEnthalpy}) to give
$h(p)$, and then inverting.  It is slightly more convenient to perform
this inversion to give $\epsilon(h)$, from which it is straightforward
to determine $p(h)$ through Eqs.~(\ref{e:p1Interpolate}) and
(\ref{e:piInterpolate}):
\begin{eqnarray}
\!\!\!\!\!\epsilon(h)&=& \epsilon_1\left\{\frac{\epsilon_1}{p_1}
\left[\exp\left(\frac{c_1-1}{c_1}\, h\right)-1\right]
\right\}^{1/(c_1-1)}
\end{eqnarray}
for $h\leq h_1$, and
\begin{eqnarray}
&&\!\!\!\!\!\epsilon(h)=\nonumber\\
&&\!\!\!\!\! \epsilon_i\left\{\frac{\epsilon_i+p_i}{p_i}
\exp\left[\frac{c_{i+1}-1}{c_{i+1}} (h-h_i)\right]-\frac{\epsilon_i}{p_i}
\right\}^{1/(c_{i+1}-1)}
\end{eqnarray}
for $h_i\leq h \leq h_{i+1}$.

\bibstyle{prd} 
\bibliography{../References/References}
\end{document}